\documentclass[aps,prb,twocolumn,floatfix,superscriptaddress,showpacs,showkeys]{revtex4}
\usepackage{graphicx}

\begin{document}

\title{Selective coherent destruction of tunneling in a quantum-dot array}
\author{J.~M.~Villas-B\^{o}as}
\affiliation{Department of Physics and Astronomy, Nanoscale and
Quantum Phenomena Institute, \\Ohio University, Athens, Ohio
45701-2979}
\affiliation{Departamento de F\'{\i}sica, Universidade
Federal de S\~{a}o Carlos, 13565-905, S\~{a}o Carlos, S\~{a}o
Paulo, Brazil}
\author{Sergio E.~Ulloa}
\affiliation{Department of Physics and Astronomy, Nanoscale and
Quantum Phenomena Institute, \\Ohio University, Athens, Ohio
45701-2979}
\author{Nelson Studart} \affiliation{Departamento de
F\'{\i}sica, Universidade Federal de S\~{a}o Carlos, 13565-905,
S\~{a}o Carlos, S\~{a}o Paulo, Brazil}
\date{\today} 

\begin{abstract}
The coherent manipulation of quantum states is one of the main
tasks required in quantum computation. In this paper we
demonstrate that it is possible to control coherently the
electronic position of a particle in a quantum-dot array. By
tuning an external ac electric field we can selectively suppress
the tunneling between dots, trapping the particle in a determined
region of the array. The problem is treated non-perturbatively by
a time-dependent Hamiltonian in the effective mass approximation
and using Floquet theory. We find that the quasienergy spectrum
exhibits crossings at certain field intensities that result in the
selective suppression of tunneling.
\end{abstract}

\maketitle

The search for a solid-state based quantum computer device has
attracted a lot of interest in the physics community recently. The
ability to manipulate a quantum state and measure it is one of the
most important and pressing challenges to be addressed in real
implementations. Fortunately great advances in this area have been
achieved recently. A good example of that are the Rabi
oscillations observed in exciton states of self-assembled quantum
dots (QDs). \cite{Stievater01,Kamada01,Htoon02,Zrenner02}

Our work here is based on the coherent destruction of tunneling
(CDT) of a driven two-level
system,\cite{Grossmann91,Bavli93,Grifoni98} in which the tunneling
of one particle in a symmetric double well potential is suppressed
for some special value of frequency and field intensity. We
propose a system in which one can selectively suppress the
tunneling between individual quantum dots tuned by the external ac
electric field. This effect provides one with a different
experimental handle to control a quantum mechanical system. By
suitable variation of the frequency and applied ac field
amplitude, one can precise the location of one electron in the
multidot array. To demonstrate this effect, our model employs an
effective mass nearest-neighbor tight-biding approximation
(NNTB).\cite{Holthaus94,Rivera00,Schulz02} The dynamics is
analyzed using Floquet theory and the direct integration of the
time-dependent Schr\"{o}dinger equation.

The Hamiltonian for an electron in an array of identical QDs under
a strong ac field within NNTB is written as
\begin{equation}
H=\sum_{j}T_{e}(a_{j+1}^{\dag}a_{j}+h.c.)+\sum_{j}eFdja_{j}^{\dag}a_{j}\cos
(\omega t+\phi ),  \label{t1}
\end{equation}%
where $T_{e}$ is the hopping matrix element, $a_{j}^{\dag}$ and
$a_{j}$ are, respectively, the electron creation and annihilation
operator in the dot $j$, $e$ is the electronic charge, $F$ is the
field intensity, $d$ is the separation between dots, $\omega$ is
the field frequency, and $\phi$ is the phase of the drive field.
This phase, first thought to be a non-relevant factor, is in fact
quite an important parameter in the CDT at smaller frequency, as
we have recently shown.\cite{Boas03} Indeed, a phase $\phi=\pi/2$
produces a much better dynamic localization than any other
possible phase $\phi$, and consequently it is our choice here.

Since $H$ is periodic in time ($H(t)=H(t+\tau)$, where $\tau =2\pi
/\omega$ is the period) we can make use of the standard Floquet
theory \cite{Boas02,Grifoni98} and write the solutions of the
time-dependent Schr\"{o}dinger equation as $\psi(t)=\exp
(-i\varepsilon t/\hbar)u(t)$, where $u(t)$, the so-called Floquet
state, is also periodic in time with the same period $\tau$, and
$\varepsilon$ is the Floquet characteristic exponent or
quasienergy, which can be obtained from the eigenvalue equation

\begin{equation}
\left( H-i\hbar \frac{\partial }{\partial t}\right)
u(t)=\varepsilon u(t). \label{t2}
\end{equation}

Note that this equation is similar to the time-independent
Schr\"{o}dinger equation with  $\mathcal{H}=H-i\hbar \partial
_{t}$ playing the role of a time-independent Hamiltonian. Using
this analogy we can explore a combined dynamic parity operation:
$z\rightarrow -z;t\rightarrow t+\tau/2$, under which the operator
$\mathcal{H}$ is invariant. As a consequence, each Floquet state
is either even or odd under this operation.\cite{Boas02}
Quasienergies of different ``dynamic parity" may cross, while an
avoided crossing is expected as a function of external parameters,
such as the field intensity, for states with the same parity.

In a symmetric double QD the levels present strictly different
parity, and they may exhibit crossings with field unless the
symmetry is broken. If we vary the field intensity, and the
quasienergy levels cross, a CDT is expected, since a splitting
between levels comes from this inter-dot tunneling. This CDT
result is well known in the high frequency limit, and it occurs at
field values satisfying the zeroes of the Bessel function,
$J_{0}(eFd/\hbar \omega)=0$. As we have shown in Ref.\
\onlinecite{Boas02}, and extended by Creffield, \cite{Creffield03}
CDT may also occur at lower frequency with decreasing degree of
localization and at different intensities of the driven field. The
dynamics of the system can be drastically different, depending on
the numbers of dots. We note that for this effect to be observed,
the frequency of the driven field cannot be excessively high,
compared with the tunneling probability between nearest dots, as
otherwise the level splitting  will be so small as to be
unnoticeable. It should also be pointed out that in the limit of
an infinite number of identical QDs the levels form a miniband
with bandwidth of $4T_{e}$, and since there is an infinite number
of levels in that range, they are infinitesimally separated, so
that even at low frequencies the levels collapse at the zeroes of
the Bessel function. Notice that the applicability of the NNTB in
that limit system has some limitations.
\cite{Domachuk02,Dignam02,Zhao94} The system considered here has
just a few QDs and we assume weak coupling between them, so that
the NNTB can be applied with confidence. Notice also that we are
considering a different kind of dynamic localization since we do
not focus only in the probability to find the particle in the same
initial state, but in a determined region of the dot array that
may not be necessarily the initial localized state. In fact the
particle stays confined to a determined multidot region of the
array.

\begin{figure}[tbp]
\includegraphics*[width=1.0\linewidth]{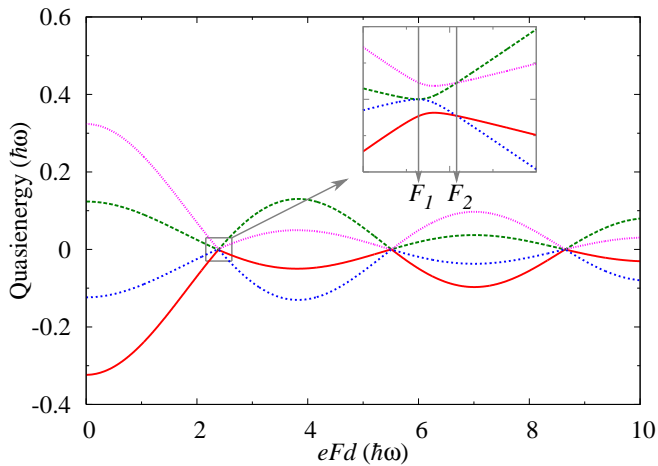}
\caption{(Color online) First Brillouin zone of quasienergies (in
units of $\hbar \protect\omega $) as function of ac field
intensity ($eFd/\hbar \protect\omega $), for $T_{e}/\hbar
\protect\omega =0.2$. Inset shows amplification of collapse
region, showing crossings and anticrossing that follow a definite
pattern according to dynamical symmetries. The two vertical lines
indicate field intensity $F_1$ and $F_2$ of pair crossings and are
used later in numerical simulation.} \label{quasi}
\end{figure}

In our analysis, we focus on the dynamics of four QDs. However, we
show that similar results can be obtained for other numbers of QDs
as well. In Fig.\ \ref{quasi} we show the quasienergy spectrum as
a function of the field intensity for the ratio $T_{e}/\hbar
\omega =0.2$ exhibiting the strong dependence on field amplitude
and the expected collapse of the spectrum near field values for
which $J_{0}(eFd/\hbar \omega)=0$. Notice however, as shown in the
inset, that the region of the level collapse is in fact a region
with several crossings and anticrossings following a well-defined
pattern given by the dynamic parity. The two vertical lines in the
inset are shown to indicate the two field values, $F_1$ and $F_2$,
where pairs of levels cross and are the choices of field intensity
used for the numerical simulations below.

\begin{figure}[tbp]
\includegraphics*[width=1.0\linewidth]{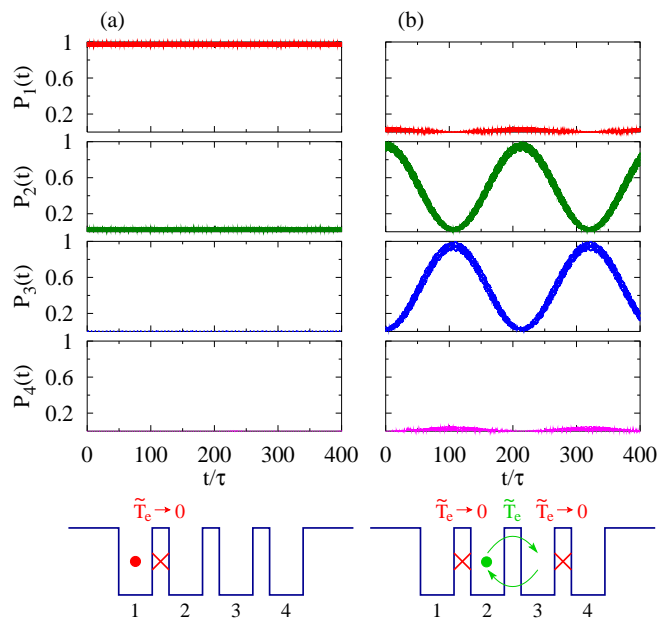}
\caption{(Color online) Time evolution of probability to find one
particle in each one of the QDs in a four-dot array for
$eF_{1}d/\hbar\omega\simeq 2.38$, and starting the system with the
particle in: (a) dot 1, and (b) dot 2. Lower panel is a schematic
representation of the dynamics of the system, as seen in the
respective time evolution of the occupation probability of each
dot. The full circle represents the position of the particle at
time zero, crosses indicate the suppression of tunneling through
that barrier and arrows indicate that tunneling is possible.}
\label{P238}
\end{figure}

Our numerical simulation is done by direct integration of the
time-dependent Schr\"{o}dinger equation and followed by
calculations of the occupation probability for different choices
of initial conditions. In Fig.\ \ref{P238}(a) we show the time
evolution of the system for the field intensity $F_1$
corresponding to the first vertical line in the inset of Fig.\
\ref{quasi} ($eF_{1}d/\hbar \omega \simeq 2.38$), assuming that at
time zero the particle is in the first QD. Notice that the
particle is basically frozen in that dot, which means that for
this choice of field we can effectively suppress the tunneling in
the barrier between dots 1 and 2. This is what the bottom cartoon
in Fig.\ \ref{P238} represents, where the cross in the barrier
indicates that the tunneling is not allowed and that the effective
interdot tunneling amplitude $\widetilde{T}_e\rightarrow0$. On the
other hand, if we start the system with the particle in dot 2,
Fig.\ \ref{P238}(b) shows that the particle can tunnel back and
forth between dots 2 and 3, while the effective tunneling is not
only suppressed between dots 1 and 2, but between 3 and 4 also.
The bottom panel represents the dynamics for this initial
condition. We can then conclude that for this first choice of
field intensity one can selectively block the tunneling of the
electron through the outer barriers, so that the only open barrier
for tunneling is the one in the middle. \footnote{From the time
evolution, one can see that $\widetilde{T}_e$ for the middle/open
barrier is $\widetilde{T}_e \simeq T_e/80$.}

\begin{figure}[tbp]
\includegraphics*[width=1.0\linewidth]{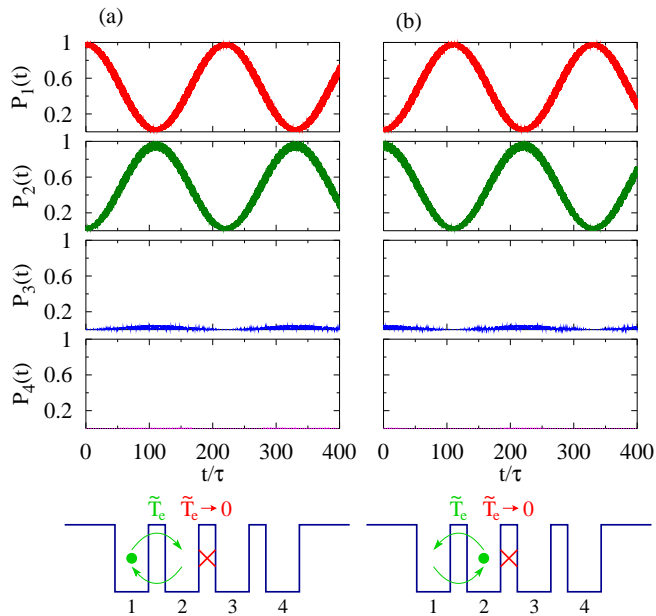}
\caption{(Color online) Time evolution of the probability to find
a particle in one of the QDs for $eFd/\hbar\omega\simeq 2.40$,
starting the system with the particle in: (a) dot 1, and (b) dot
2. Lower panel is a schematic representation of the dynamics.
Here, the middle barrier tunneling is suppressed.} \label{P240}
\end{figure}

Let us now explore the second choice of field intensity $F_2$, the
second vertical line in the inset of Fig.\ \ref{quasi},
($eF_2d/\hbar \omega \simeq 2.4$). This small but discernible
tuning of the field intensity yields a completely different
result. At that point there are two crossings in the spectrum. In
Fig.\ \ref{P240}(a), we show the results when we start the system
with the particle in dot 1, and in Fig.\ \ref{P240}(b) starting in
dot 2. The picture for both initial conditions does not change at
all, except for a $\pi$-phase in the oscillation (that is provided
by the initial state). The effective tunnelings can be summarized
in the respective bottom panels of that figure, and we can see
that now we can block the tunneling between dots 2 and 3. Similar
behavior and conclusions are reached for initial conditions in
either dot 3 or 4.

\begin{figure}[tbp]
\includegraphics*[width=1.0\linewidth]{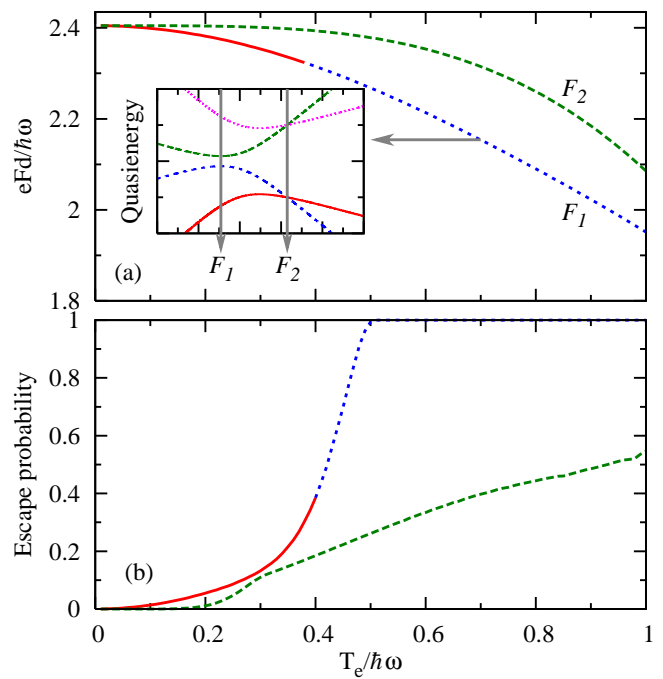}
\caption{(Color online) (a) Crossing in the collapse region as
indicated by vertical lines in the inset of Fig.\ \ref{quasi}, as
function of the ratio $T_{e}/\hbar \omega$. In the high frequency
limit (lower ratio $T_{e}/\hbar \omega$) the crossings occur
basically at the same point (zeroes of Bessel function). Inset
shows amplification of quasienergy spectrum for $T_{e}/\hbar
\omega=0.7$. (b) Escape probability for a particle initially in
one region of the quantum-dot array. Solid red line is for the
equivalent first crossing and represents the escape probability of
the QD 1, considering this dot as the initial condition. Dotted
blue line continuation shows $F_1$ when the crossing disappears.
Dashed green line is for the second crossing and represents escape
probability of the first two QDs assuming that the system starts
in QD 1. } \label{localization}
\end{figure}

These examples represent an exciting result since by simply tuning
the field intensity from $F_1$ to $F_2$, one can choose which
barrier is allowed or blocked for tunneling. Notice that this dot
or site selectivity is achieved despite the ac field being applied
to the entire structure, which suggests interesting applications.
Notice, moreover, that the amount of field intensity needed to
change this condition is relative to the frequency of the driving
field, i.e., the ratio $T_{e}/\hbar\omega$. It is also interesting
that the separation between the two tunneling suppressing fields
is not a monotonic function of $T_{e}/\hbar\omega$. Figure
\ref{localization}(a) shows how these two crossing points change
with the ratio $T_{e}/\hbar \omega$. The solid line shows the
first crossing at field intensity $F_1$, and the dashed line is
the result for $F_2$. Notice that in the high frequency regime
(smaller ratio $T_{e}/\hbar \omega$) the crossings occur very
close to each other and at field intensities satisfying the
condition $J_0(eFd/\hbar \omega)=0$. As a result, all barriers
present a suppression of tunneling at the same value of field
intensity in the high frequency regime.\cite{Holthaus93} For lower
frequencies and given $T_e$, the crossings become separated and
different barriers are selectively closed at different fields. At
even lower frequency $T_{e}/\hbar \omega \gtrsim 0.4$, the first
crossing disappears (indicated by the dotted blue line
continuation) as the coupling between quasienergy levels one and
tree, and two and four increases. This results in an effective gap
in the quasienergy spectrum (see inset of Fig.\
\ref{localization}(a) for $T_{e}/\hbar \omega = 0.7$) that
basically kills the CDT for that value of field intensity and
frequency.

To better understand how good is this CDT at values of field given
by $F_1$ and $F_2$ we can monitor the maximum probability for the
particle tunneling out of the region of dots we are considering
(see Ref.\ \onlinecite{Boas02}). Lower values for this represent a
good CDT. Figure \ref{localization}(b) shows the escape
probability for field intensities $F_1$ and $F_2$ provided by the
crossings in Fig.\ \ref{localization}(a), as function of the ratio
$T_{e}/\hbar \omega$ (frequency). The solid line is the result for
the first crossing $F_1$, and is the escape probability out of dot
1, defined as the maximum value reached by $P_2+P_3+P_4$, assuming
that the system starts with the particle in dot 1. The same result
is clearly obtained if the system starts with the particle in dot
2 and one monitors the maximum probability to find it in dot 1.
The dotted blue line that follows the solid red line in Fig.\
\ref{localization}(b) is the result for field values $F_1$ at
frequencies when there are no more crossing, but a gap in the
quasienergy spectrum. Notice this goes to unity quickly as the
effective tunneling becomes possible as a result of having a gap
in the quasienergy spectrum. The dashed green line in Fig.\
\ref{localization}(b) is the result for the second level crossing
field $F_2$, representing the escape probability of the dots 1 and
2 (since for this choice of field we suppress the tunneling
between dots 2 and 3), which is the maximum value reached by
$P_3+P_4$ starting the system with the particle in either dots 1
or 2. Notice that for frequencies around $T_{e}/\hbar \omega
\simeq 0.3$ the system still presents a good CDT for both
crossings, and they occurs at field intensity with significant
separation.

\begin{figure}[t]
\includegraphics*[width=1.0\linewidth]{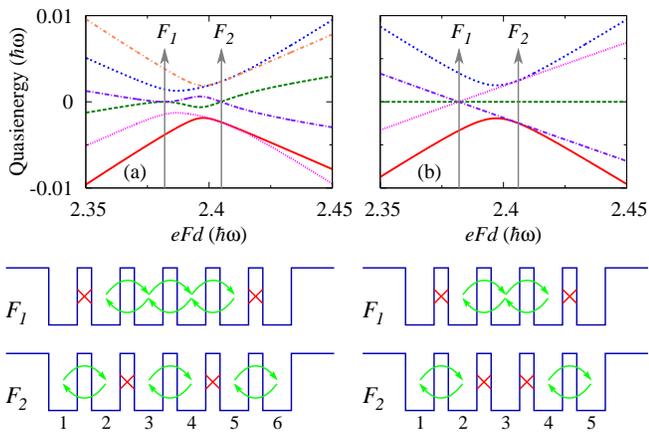}
\caption{(Color online) Collapse region of the first Brillouin
zone of quasienergies (in units of $\hbar \protect\omega$) as
function of the ac field intensity ($eFd/\hbar \protect\omega$),
for $T_{e}/\hbar \protect\omega =0.2$ for: (a) an array of six QDs
and (b) an array of five QDs. Lower cartoons are the schematic
representation of the dynamic of the respective arrays of QDs at
field intensity at crossing points given by $F_1$ and $F_2$. }
\label{qd65}
\end{figure}

Our analysis was given for four QDs, but the same behavior can be
obtained for other finite number of dots. The quasienergy spectrum
presents two crossings, $F_1$ and $F_2$, in the collapse region as
we can see in Fig.\ \ref{qd65}(a) for six and (b) for five QDs,
respectively. The first crossing $F_1$ basically suppresses the
tunneling between dots 1 and 2, and between the two last QDs. This
result is schematically represented in the bottom cartoons for
$F_1$ for either five or six QDs. The second crossing in the
spectrum, $F_2$, suppress the tunneling between pairs of quantum
dots, for example, in the six QDs cases, the tunneling between
dots 2 and 3, and 4 and 5 as schematically represented in the
lower bottom panel of Fig.\ \ref{qd65}(a). An interesting result
appears for odd numbers of QDs, since in this case there is a
different number of even and odd states. This fact allows for the
dot in the middle to become a true trap for the particle, since
tunneling in both directions can be suppressed. For example, the
case of five QDs for the second crossing, $F_2$, as schematically
represented by the lower bottom cartoon of Fig.\ \ref{qd65}(d),
suppress the tunneling between the dots 2 and 3, and 3 and 4. So,
if we could start the system, or manipulate it, in a state
localized in that dot and choose this value of field intensity
($F_2$) we could in principle trap the particle in the middle of
the quantum-dot array.

In conclusion, we have shown that it is possible to suppress
selectively the tunneling between quantum dots by a simple tuning
from $F_1$ to $F_2$, the intensity of an applied ac field. With
this tool one could manipulate the position of the particle in a
quantum-dot array, and assist in its initialization and control.

This work was partially supported by FAPESP, US DOE grant no.
DE-FG02-91ER45334 and the CMSS Program at Ohio University.


\begin{thebibliography}{9}


\bibitem{Stievater01}T. H. Stievater, Xiaoqin Li, D. G. Steel, D.
Gammon, D. S. Katzer, D. Park, C. Piermarocchi, and L. J. Sham,
Phys. Rev. Lett. \textbf{87}, 133603 (2001).

\bibitem{Kamada01}H. Kamada, H. Gotoh, J. Temmyo, T. Takagahara,
and H. Ando, Phys. Rev. Lett. \textbf{87}, 246401 (2001).

\bibitem{Htoon02}H. Htoon, T. Takagahara, D. Kulik, O. Baklenov,
A. L. Holmes Jr., and C. K. Shih, Phys. Rev. Lett. \textbf{88}, 087401 (2002).

\bibitem{Zrenner02}A. Zrenner, E. Beham, S. Stufler, F. Findeis,
M. Bichler, and G. Abstreiter, Nature (London) \textbf{418}, 612 (2002).

\bibitem{Grossmann91} F. Grossmann, T. Dittrich, P. Jung, and P. H\"anggi, Phys. Rev.
Lett. \textbf{67}, 516 (1991).

\bibitem{Bavli93} R. Bavli and H. Metiu, Phys. Rev. A \textbf{47}, 3299 (1993).

\bibitem{Grifoni98} M. Grifoni and P. H\"{a}nggi, Phys. Rep. \textbf{304}, 229 (1998).

\bibitem{Holthaus94} M. Holthaus and D. W. Hone, Phys. Rev. B \textbf{49}, 16605 (1994).

\bibitem{Rivera00} P. H. Rivera and P. A. Schulz, Phys. Rev. B \textbf{61}, R7865 (2000).

\bibitem{Schulz02} P. A. Schulz, P. H. Rivera, and N. Studart Phys. Rev. B \textbf{66}, 195310 (2002)

\bibitem{Boas03} J. M. Villas-B\^{o}as, S. E. Ulloa, and N. Studart, unpublished.

\bibitem{Boas02} J. M. Villas-B\^{o}as, W. Zhang, S. E. Ulloa, P.
H. Rivera, and N. Studart, Phys. Rev. B \textbf{66}, 085325
(2002).

\bibitem{Creffield03} C. E. Creffield, Phys. Rev. B \textbf{67},
165301 (2003).

\bibitem{Domachuk02} P. Domachuk, C. M. de Sterke, J. Wan,
and M. M. Dignam, Phys. Rev. B \textbf{66}, 165313 (2002).

\bibitem{Dignam02} M. M. Dignam and C. M. de Sterke,
Phys. Rev. Lett. \textbf{88}, 046806 (2002).

\bibitem{Zhao94} X.-G. Zhao, J. Phys. Condens. Matter \textbf{6}, 2751 (1994).

\bibitem{Holthaus93} M. Holthaus and D. Hone, Phys. Rev. B \textbf{47}, 6499 (1993)

\end{thebibliography}
\end{document}